\documentclass[twocolumn,showpacs,preprintnumbers,amsmath,amssymb]{revtex4}

\usepackage{graphicx}
\usepackage{dcolumn}
\usepackage{bm}
\usepackage{textcomp}
\usepackage{wasysym}
\usepackage{textcomp}
\usepackage{mathrsfs}
\usepackage{mathrsfs}

\begin{document}
\title{Equal-Interval Splitting of Quantum Tunneling in Single-Molecule Magnets with Identical Exchange Coupling}
\author{Yan-Rong Li,  Rui-Yuan Liu, Hai-Qing Liu and Yun-Ping Wang}
\affiliation{Beijing National Laboratory for Condensed Matter
Physics,  Institute of Physics, Chinese Academy of Sciences, Beijing
100190, People's Republic of China}


\date{\today}
\begin{abstract}
The equal-interval splitting of quantum tunneling observed in
simple-Ising-model systems of Ni$_{4}$ (3D) and Mn$_3$ (2D)
single-molecule magnets (SMMs) is reported. The splitting is due to
the identical exchange coupling in the SMMs, and is simply
determined by the difference between the two numbers of the
spin-down $n_{\downarrow}$ and spin-up $n_{\uparrow}$ molecules
neighboring to the tunneling molecule. The splitting may be
presented as $(n_{\downarrow}-n_{\uparrow})JS/{g\mu_{0}\mu_{B}}$,
and the number of the splittings follows $n+1$  where
$n=n_{\downarrow}+n_{\uparrow}$  is the coordination number.
Besides, since the quantum tunneling is heavily dependent on  local
spin environment, the manipulation of quantum tunneling may become
feasible for this kind of system,  which may shed new light on novel
applications of SMMs.
\end{abstract}
\maketitle

Single-molecule magnets (SMMs) have been used as model systems to
study the interface between classical and quantum behaviors, and are
considered to be the most promising systems for the applications in
quantum computing, high-density information storage and magnetic
refrigeration \cite{computing,qubit,Spintronics,refriPoly,refriAPL}
due to the quantum tunneling of magnetization (QTM) observed in
these systems \cite{Mn12Nature,Mn12PRL,FePRL,FeJACS}. Recent
researches in the impact of intermolecular exchange couplings upon
the QTM have focused on whether the exchange coupling may change the
quantum tunneling in SMMs. SMM dimer system is reported to have
different quantum behavior from that of the individual SMMs, due to
the intermolecular exchange couplings between the two components
\cite{dimerNature,dimerPRL}. It is also reported that, in the SMM
dimer with 3D network of exchange-couplings, the QTM is not
suppressed \cite{dimerPRB}. In this letter, we demonstrate that, for
the SMMs with identical exchange coupling(IEC), the quantum
tunneling behavior is much simpler and the QTM might be conveniently
manipulated by controlling of the magnetization.

In the following, we report a unique quantum tunneling effect
observed in the single-molecule magnets of
[Ni(hmp)(CH$_{3}$CH$_{2}$OH)Cl]$_{4}$ (hereafter Ni$_{4}$)
\cite{Ni4bias,Ni4fast} and
[Mn${_3}$O(Et-sao)${_3}$(MeOH)${_3}$(ClO${_4}$)] (hereafter
Mn$_{3}$) \cite{Mn3distort,Mn3FM}. Ni$_{4}$ SMM  is a crystal with
3D network of exchange coupling, in which each molecule is coupled
with four neighboring molecules by Cl$\cdot\cdot\cdot$Cl contact
(which contributes to the exchange coupling) forming a diamond-like
lattice. Ni$_{4}$ crystal has $S_4$ symmetry, which ensure that the
four exchange couplings between each molecule and its four
neighboring molecules are identical throughout the crystal. Mn$_3$
SMM is a crystal with 2D network of exchange coupling, in which each
molecule is coupled with three neighboring molecules by hydrogen
bonds (which contributes to the exchange coupling) in ab plain,
forming a honeycomb-like structure viewed down along the c-axis.
Mn$_3$ crystal has $C_3$ symmetry, which ensure that the three
exchange couplings between each molecule and its three neighboring
molecules are identical throughout the crystal. We notice that both
Ni$_{4}$ and Mn$_3$ SMMs are crystals with IEC and the model systems
of simple Ising model \cite{Ising}. We have observed the
equal-interval splitting of quantum tunneling induced by IEC in
these two systems by ac susceptibility and hysteresis loop
measurements.

Considering the low blocking temperature, we studied quantum
tunneling effects of Ni$_4$ SMM by ac susceptibility measurements,
with a home-made compensation measurement setup \cite{Li}.  Fig.1
has demonstrated the temperature dependence of the quantum tunneling
behavior in Ni$_4$ SMM. Apparently, the peak at zero field
disappears at 0.75K and 0.5K, which consists with the missing step
at zero field in magnetization hysteresis loops at 40mK
\cite{Ni4bias}. As a result of different orientations, the step
positions are different from those mentioned in Ref\cite{Ni4bias}.
We measured the quantum tunnelings at different orientations and
\begin{figure}[hb] \scalebox{0.38}{\includegraphics[bb=400 50 8cm
16.5cm]{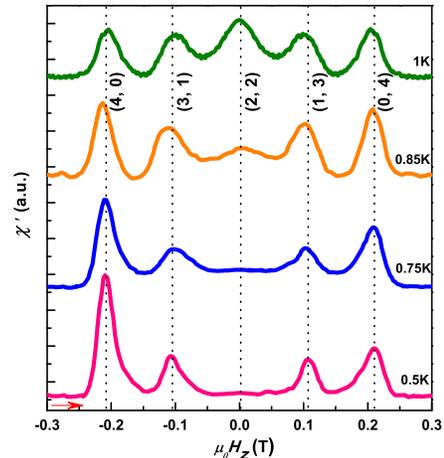}}
 \caption
{(Color online). Field dependence of susceptibility $\chi{'}$ from
$-$0.3T to 0.3T at different temperatures measured on Ni$_4$ single
crystal, with the sweeping rate of 0.001T/s. The quantum tunneling
peaks marked by black dotted lines origin from tunnelings between
$|-4\rangle$ and $|4\rangle$ spin state, and the labeled number set
in () besides each dotted line indicates the local spin environment
$(n_{\downarrow},n_{\uparrow})$ of the tunneling molecules. }
\end{figure}
found the resonant fields along the easy axis of the sample are
$-0.21$T $-0.11$T, 0T, 0.11T, 0.21T as shown in Fig.1. It is seen
that the tunneling peaks appear with equal interval. The shift of
the tunneling peaks from higher to lower field  with the increasing
T is due to the enhancement of the effect of  thermal activation
upon tunneling \cite{cross,selection}.

The higher blocking temperature allows us to study the hysteresis
loops above 1.6K  for Mn$_3$ SMM. Fig.2 shows the  typical step-like
hysteresis loops of Mn3 SMM at different temperatures.  The blocking
temperature estimated from ZFC (zero field cooling) and  FC (field
cooling) curves shown in the  inset is around 3K.   The
sweep-rate-dependent magnetization curves at 1.6K are shown in
Fig.3,  with only a dM/dH curve at the sweeping rate of 0.0005T/s
presented for simplicity. A series of quantum tunneling peaks with
an equal interval of 0.36T are observed in the dM/dH curves, which
is similar to those observed in Ni$_4$ SMM.

With IEC taken into account, the molecules are not isolated, and the
spin Hamiltonian  of each molecule may be presented as:
\begin{equation}
\hat{\mathcal{H}}=-D\hat{S}^2_z +
g\mu_{0}\mu_{B}\hat{S}_zH_z-\sum_{i=1}^{n}J\hat{S}_{z}\hat{S}_{iz},
\end{equation}
where $D$ is the axial anisotropy constant, $n$ is coordination
number, $J$ is the exchange interaction constant,  $\hat{S}_{z}$ and
$\hat{S}_{iz}$ are the easy-axis spin operators of the molecule and
its $i$th exchange-coupled neighboring molecule. For Ni${_4}$,
$S=4$, $D=0.86$K, $g=2{_\cdot12}$ \cite{Ni4bias,Ni4fast}; while for
Mn${_3}$, $S=6$, $D=0.98$K, $g=2{_\cdot06}$ \cite{Mn3FM}.

In Ni$_4$ SMM, every Ni$_4$ molecule has four AFM exchange-coupled
neighboring molecules, and hence for each molecule there are five
different kinds of local spin environment (LSE), which may be
labeled by $(n_{\downarrow},n_{\uparrow} )$, where $n_{\downarrow}$
 and $n_{\uparrow}$ represent the number of the neighboring molecules
 which occupy $S_{z}=-4$ (hereafter $|-4\rangle$) and $S_{z}=4$ (hereafter
$|4\rangle$) spin states respectively (The excited spin states are
not considered here,
\begin{figure}[hb]
\scalebox{0.34}{\includegraphics[bb=550 50 8cm 17cm]{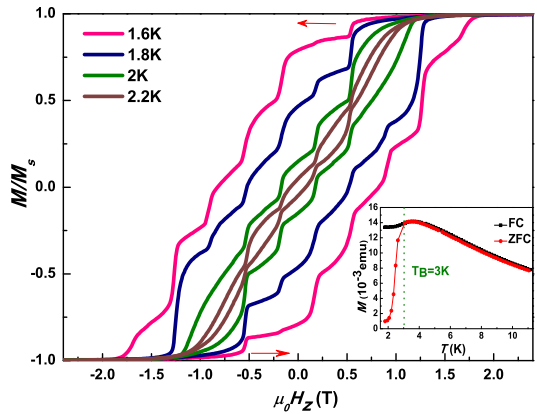}}
 \caption
{(Color online). Magnetization (M/Ms) of Mn$_3$ single crystal
versus applied magnetic field with the sweeping rate of 0.003T/s at
different temperatures. The inset shows ZFC and FC curves.}
\end{figure}
because most of them are not populated at our measurement
temperatures). At negative saturated field, all the molecules
initially occupy $|-4\rangle$ in the same LSE $(4, 0)$ shown in
Fig.4a (left). According to equation(1), $|-4\rangle$ and
$|4\rangle$ spin states in the LSE $(4, 0)$ are degenerate when the
field reaches $4JS/{g\mu_{0}\mu_B}$, therefore those molecules which
occupy the $|-4\rangle$ spin state in the LSE (4, 0) (Fig.4a) have
the same probability to undergo tunneling at $4JS/{g\mu_{0}\mu_B}$,
leading to the resonant tunneling peaks at $-$0.21T as seen in
Fig.1. Following this resonant quantum tunneling, some molecules
will occupy $|4\rangle$ spin state, and the LSE of the molecules
will not be identical any more. When the field reaches
$2JS/{g\mu_{0}\mu_B}$ (corresponding to $-$0.11T as seen in Fig.1),
the resonant tunneling takes place from $|-4\rangle$ to $|4\rangle$
spin state in the LSE (3, 1) (Fig.4b). As a matter of fact, at zero
field the tunneling of the molecules in the LSE (2, 2) (Fig.4c) will
change neither Zeeman energy nor the exchange interaction energy,
which gives rise to the macroscopic quantum tunneling  observed at
zero field at relatively higher temperatures shown in Fig.1.
Similarly, there are quantum tunnelings taking placing from
$|-4\rangle$ to $|4\rangle$ spin state with the LSE (1, 3) at
$-2JS/{g\mu_{0}\mu_B}$ , and from $|-4\rangle$ to $|4\rangle$ spin
state with the LSE (0, 4) at $-4JS/{g\mu_{0}\mu_B}$. The exchange
interaction constant J is calculated to be $-0.019$K according to
the splitting interval, which is close to the simulation value
$-0.02$K obtained from the experimental AFM transition temperature
of $T_{N}=0.91$K \cite{Zuo}. At temperatures obviously below
$T_{N}$,  the spins of the molecules will be anti-parallel to its
neighbors, i.e. the molecules are in the LSE (0, 4) and (4, 0)
instead of the LSE (2, 2), which causes the missing of quantum
tunneling at zero field at
\begin{figure}[ht]
\scalebox{0.38}{\includegraphics[bb=550 50 8cm 18.5cm]{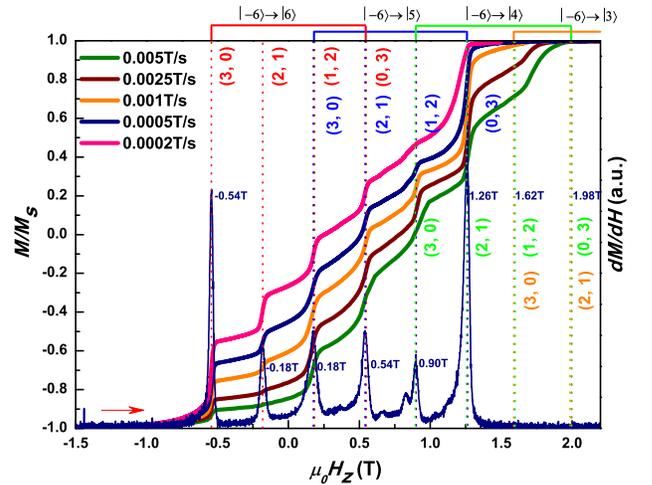}}
 \caption
{(Color online). Sweep-rate-dependent magnetization (M/M$_s$) versus
applied magnetic field $\mu_{0}$H$_z$ (from $-$1.5T to 2.2T) at 1.6K
measured on Mn$_3$ single crystal. dM/dH curve with sweeping rate of
0.0005T/s is given. The quantum tunneling peaks marked by the dotted
lines of the same color belong to the same tunneling of
${|m_{i}\rangle\rightarrow|m_{f}\rangle}$, the labeled number set in
() besides each dotted line indicates the local spin environment
$(n_{\downarrow},n_{\uparrow})$ of the tunneling molecules. }
\end{figure}
$T\leq0.75$K as seen in Fig.1. However, in the vicinity of  the
transition temperature, some molecules are still in the LSE (2, 2)
due to the thermal fluctuation, thus there is still an evidence of
resonant quantum tunneling at 0.85K at zero field shown in Fig.1.

\begin{figure}[hb] \scalebox{0.4}{\includegraphics[bb=180 20
10cm 15cm]{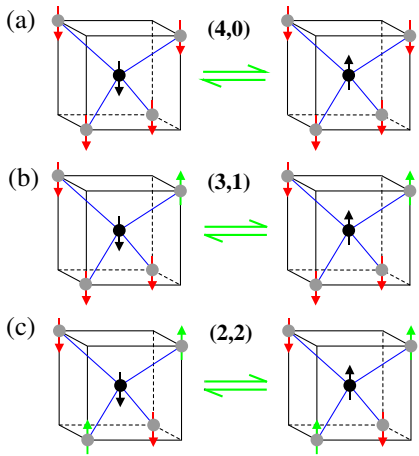}}
 \caption
{(Color online). Sketch maps of three pairs of spin configurations
with different LSE $(n_{\downarrow},n_{\uparrow})$ in Ni$_4$ SMM, in
correspondence to the tunnelings occurring at $-$0.21T, $-$0.11T and
0T in Fig.1, respectively. Other equivalent spin configurations are
not listed here for simplicity. The tunneling molecule is marked in
black with black arrow indicating its spin state, its four neighbors
are marked in gray, with green and red arrows indicating spin-up and
spin-down state respectively, the blue lines between molecules
represents the exchange couplings.}
\end{figure}
Mn$_3$ SMM displays a finer quantum tunneling behavior than Ni$_4$
SMM.  Every Mn$_3$ molecule has three AFM exchange-coupled
neighboring molecules, and hence for each molecule there are four
different kinds of local spin environment as shown in Fig.5, labeled
as (3, 0), (2, 1), (1, 2), (0, 3) respectively, thus, there are
quantum tunnelings occurring at $3JS/{g\mu_{0}\mu_{B}}$,
$JS/{g\mu_{0}\mu_{B}}$, $-JS/{g\mu_{0}\mu_{B}}$,
$-3JS/{g\mu_{0}\mu_{B}}$ from $|-6\rangle$ to $|6\rangle$ spin
state, which is corresponding to the four tunneling peaks marked by
the red dotted lines shown in Fig.3.

In both Mn$_3$ and Ni$_4$ SMMs, with the presence of IEC, the
tunneling between two ground spin states of $|\pm S\rangle$ is
splitted by equal-interval field of $2|J|S/{g\mu_{0}\mu_{B}}$.
Generally, according to equation(1), the tunneling from $|-S\rangle$
to $|S-l\rangle$ is splitted by the same equal-interval field, and
the splitted tunneling field may be simply expressed as
\begin{equation}
H_z=lD/g\mu_{0}\mu_{B}+(n_{\downarrow}-n_{\uparrow})JS/{g\mu_{0}\mu_{B}}.
\end{equation}
The first term comes from the internal spin states in each molecule,
and the second term is of the tunneling splitting induced by IEC.
The splitting is simply determined by the difference between the two
numbers of the spin-down( $n_{\downarrow}$) and spin-up
($n_{\uparrow}$) molecules neighboring to the tunneling molecule.
According to equation(2), the number of splittings equals the number
of different kinds of ($n_{\downarrow}$, $n_{\uparrow}$) LSEs, and
hence may be expressed as ${n+1 \choose 1}=n+1$ by combinatorics,
where $n = n_{\downarrow} + n_{\uparrow}$.

According to equation(2), when $D>n|J|S$ and
$|H_z|<(D-n|J|S)/g\mu_{0}\mu_{B}$, any quantum tunneling with $l
\neq 0$ are not allowed; while according to equation(1), when the
first excitation energy of a molecule $D(2S-1)\gg kT$, almost all
molecules will occupy the two ground spin states of $|\pm S\rangle$.
Therefore, under the above conditions, equation(1) may be simplified
as
\begin{equation}
\hat{\mathcal{H}}=g\mu_{0}\mu_{B}\hat{S}_zH_z-\sum_{i=1}^{n}J\hat{S}_{z}\hat{S}_{iz},
\end{equation}
which is just the Hamiltonian of simple Ising model \cite{Ising}.
For both Ni${_4}$ and Mn${_3}$, $D>n|J|S$, thus Ni${_4}$ and
Mn${_3}$ SMMs are good model systems of simple Ising model at low
temperature and low field, which are important for the studies of
quantum tunneling behavior and related applications.

\begin{figure}[hb] \scalebox{0.15}{\includegraphics[bb=850 20 10cm
36cm]{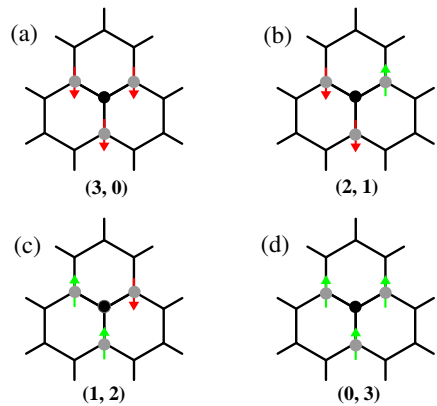}}
 \caption
{(Color online). Sketch maps of four spin configurations with
different LSE $(n_{\downarrow},n_{\uparrow})$ in Mn$_3$ SMM, other
equivalent spin configurations are not listed here for simplicity.
The tunneling molecule is marked in black, its three neighboring
molecules are marked in grey, with green and red arrows indicating
spin-up and spin-down state respectively. The black lines between
molecules represent the exchange couplings. }
\end{figure}
Since the intermolecular exchange couplings are identical in the
system, the magnitude $\mathcal{T}$ of a tunneling may be simply
factorized into  intermolecular contribution
$N_{(n_{\downarrow},n_{\uparrow})}$ and  intramolecular contribution
$P_{|m_{i}\rangle\rightarrow|m_{f}\rangle}$,
\begin{equation}
\mathcal{T}=\alpha N_{(n_{\downarrow},n_{\uparrow})}
P_{|m_{i}\rangle\rightarrow|m_{f}\rangle},
\end{equation}
where $N_{(n_{\downarrow},n_{\uparrow})}$ is the number of molecules
with the LSE $(n_{\downarrow},n_{\uparrow})$,
$P_{|m_{i}\rangle\rightarrow|m_{f}\rangle}$ is the tunneling
probability of the molecule from the spin state $|m_{i}\rangle$ to
$|m_{f}\rangle$, and $\alpha$ is a constant.
$N_{(n_{\downarrow},n_{\uparrow})}$ strongly depends on the
magnetization M and may be easily modulated, while
$P_{|m_{i}\rangle\rightarrow|m_{f}\rangle}$ is determined by the
tunneling barrier between $|m_{i}\rangle$ and $|m_{f}\rangle$ inside
molecules and is hardly to be controlled. Therefore, for
SMMs-with-IEC, with the dependence of $\mathcal{T}$ on
$N_{(n_{\downarrow},n_{\uparrow})}$,  the manipulation of quantum
tunneling should be rather simple.

The quantum tunnelings from the same initial states $|m_{i}\rangle$
to the same final states $|m_{f}\rangle$ but with different LSEs are
referred to as a tunneling set. The five tunneling peaks of Ni$_4$
SMM in Fig.1, belong to the same set of
$|-4\rangle\rightarrow|4\rangle$ and has the same
$P_{|m_{i}\rangle\rightarrow|m_{f}\rangle}$, thus the intensities of
the five peaks is proportional to
$N_{(n_{\downarrow},n_{\uparrow})}$, which means that
$N_{(n_{\downarrow},n_{\uparrow})}$ may be monitored by macroscopic
measurements of the tunneling peaks. For Mn$_3$ SMM, the AFM
exchange coupling constant J is calculated to be $J=-0.041$K
according to the field interval of the
$|-6\rangle\rightarrow|6\rangle$ tunneling set(Fig.3). However the
axial anisotropy constant $D=0.98$K \cite{Mn3FM} of Mn$_3$ SMM
happens to be close to $4|J|S$, which results in the overlap of two
adjacent tunneling sets demonstrated by the overlapped dotted lines
shown in Fig.3. The tunneling steps at 0.18T and 0.54T are the
combinations of the tunnelings from $|-6\rangle$ to $|6\rangle$ spin
state with the LSEs (1, 2) and (0, 3) (marked by red dotted lines)
and quantum tunnelings from $|-6\rangle$ to $|5\rangle$ spin state
with the LSEs (3, 0) and (2, 1) (marked by blue dotted lines)
respectively. Similarly, all subsequent tunneling steps are
combinations of quantum tunnelings in different tunneling sets with
different local spin environments. It may be worth a mention that
the tunnelings are expected to occur at 1.62T and 1.98T (marked by
green and orange dotted lines) at lower temperatures as well,
although not observed in these curves.

Of the overlapped tunnelings mentioned above, due to the dependence
of tunneling on the local spin environment, the contribution of the
individual tunneling changes as the field sweeping rate varies. For
example, the tunneling step at 0.18T is the combination of tunneling
from $|-6\rangle$ to $|6\rangle$ spin state with the LSE (1, 2) and
tunneling from $|-6\rangle$ to $|5\rangle$ spin state with the LSE
(3, 0), therefore the tunneling magnitude is determined by $N_{(3,
0)}P_{|-6\rangle\rightarrow|5\rangle}+N_{(1,
2)}P_{|-6\rangle\rightarrow|6\rangle}$, where $N_{(3, 0)}$ and
$N_{(1, 2)}$ strongly depends on the magnetization M.  As shown in
Fig.3, for the tunneling at 0.18T, M/M$_s$ is increasing with the
decreasing of field sweeping rate, which suggests that $N_{(1, 2)}$
is increasing while $N_{(3, 0)}$ is decreasing, and hence the
contribution of the tunneling from $|-6\rangle$ to $|6\rangle$ spin
state with the LSE (1, 2) is taking the dominance from the
contribution of the tunneling from $|-6\rangle$ to $|5\rangle$ spin
state with the LSE (3, 0), eventually.

Due to the strong dependency of a tunneling on the
$N_{(n_{\downarrow},n_{\uparrow})}$ based on equation(4), the
subsequent quantum tunneling heavily depends on the the preceding
quantum tunnelings in SMMs-with-IEC. As shown in Fig.3, tunneling at
$-$0.54T (from $|-6\rangle$ to $|6\rangle$ spin state with the LSE
(3, 0)) is inherited by tunneling at $-$0.18T (from $|-6\rangle$ to
$|6\rangle$ spin state with the LSE (2, 1)), the tunnelings at
$-$0.54T, $-$0.18T are further carried on by the next tunneling, and
the process continues as the LSE changes. In fact, the history
dependence is not prominent for Ni$_4$ SMM, due to that the
measurements were performed at temperatures much higher than its
blocking temperature, while thermal activated effect ruin the memory
of history. Apparently, the subsequent quantum tunneling is more
heavily dependent on the preceding quantum tunnelings in
SMMs-with-IEC when the thermal activated effect is severely
suppressed as the temperature drops adequately. This indicates a new
way for manipulating quantum tunneling.

In summary, we performed detailed ac susceptibility and hysteresis
loop measurements on Ni$_4$ and Mn$_3$ single crystals,
respectively, and have observed the equal-interval splitting of
quantum tunneling in both systems, the splitting of quantum
tunneling is presented by
$(n_{\downarrow}-n_{\uparrow})JS/{g\mu_{0}\mu_{B}}$; and the number
of splitting follows $n+1$, where $n=n_{\downarrow}+n_{\uparrow}$ is
the coordination number. Since the splitting is induced by the IEC
between the molecules, the rules should be universally applicable to
all single-molecule magnets with IEC. Besides, it is demonstrated
that, the manipulation of quantum tunneling may become feasible for
this kind of system, which may shed new light on novel applications
of SMMs.

We thank Prof. Dianlin Zhang, Lu Yu, and Li Lu for helpful
discussions. We also thank Shaokui Su for experimental assistance.
This work was supported by the National Key Basic Research Program
of China (No.2011CB921702) and the Natural Science Foundation of
China (No.11104331).

\end{document}